\documentclass[12pt]{iopart}

\expandafter\let\csname equation*\endcsname\relax
\expandafter\let\csname endequation*\endcsname\relax
\usepackage{amsmath, amssymb, color, graphicx, cases, stmaryrd, wasysym, amsbsy}

\definecolor{linkcolor}{rgb}{0,0,0.6} 
\usepackage[colorlinks=true,
	pdfstartview = FitV,
	linkcolor    = linkcolor,
	citecolor    = linkcolor,
	urlcolor     = linkcolor,
	hyperindex   = true,
	hyperfigures = false]{hyperref}

\usepackage{epsfig}
\usepackage{textcomp}
\usepackage{float}
\usepackage{braket}
\usepackage[normalem]{ulem}
\usepackage{enumerate}
\usepackage{enumitem}
\usepackage{cite}

\begin{document}

\title{Collective deformation of anisotropic particles with internal pulsation}

\author{Luca Casagrande}
\address{Department of Physics and Materials Science, University of Luxembourg, L-1511 Luxembourg, Luxembourg}

\author{Alessandro Manacorda}
\address{CNR Institute of Complex Systems, Uos Sapienza, Piazzale A. Moro 5, 00185 Rome, Italy}
\address{Department of Physics and Materials Science, University of Luxembourg, L-1511 Luxembourg, Luxembourg}

\author{\'Etienne Fodor}
\address{Department of Physics and Materials Science, University of Luxembourg, L-1511 Luxembourg, Luxembourg}

\begin{abstract}
Capturing the emergence of deformation waves in contractile living tissues is a challenge that has recently been tackled with models of actively deformable particles. Inspired by the anisotropic deformation of cardiomyocytes in cardiac tissues, we examine how the pulsation of elliptical particles affects their collective properties in dense assemblies. We introduce two types of deformation where the eccentricity of each particle is subject to a periodic drive, and examine the interplay between nematic order and synchronized deformation via a systematic phase diagram. We derive a hydrodynamic description through a coarse-graining procedure, and show that it qualitatively captures the main collective states of the microscopic dynamics. Overall, our model provides key insights into how an active anisotropic deformation yields waves that self-organize into various dynamical patterns.
\end{abstract}

\maketitle


\section{Introduction}\label{sec:intro}

Active matter refers to systems composed of units that absorb energy from their environment to sustain their motion and deformation~\cite{hydro_marchetti, fodor_lecture}. Over the past few decades, the field of active matter has expanded significantly, leading to the development of various models that describe diverse many-body systems and their collective dynamics. These include crowds of pedestrians~\cite{active_pedestrians, gu2025emergence}, bacterial colonies~\cite{DiLeonardo10pnas,bacterial_colonies,Aranson22rpp}, and biological tissues~\cite{selfpropvert1, selfpropvert2, czajkowski2018}. These studies have led to reveal intriguing collective behaviors such as flocking~\cite{vicsek2,birds} and motility-induced phase separation~\cite{selfprop1}. The interest in these phenomena is mostly rooted in their intrinsic non-equilibrium nature: by evading equilibrium constraints, active matter exhibits collective behaviors without any equilibrium equivalent.

Dense active systems have received increasing attention in the last decade. Experimental studies on confluent living tissues have revealed the important role of nonequilibrium collective behaviors of cells in various biological functions; for instance, in contexts like embryonic development~\cite{wave_obs6}, uterine contraction~\cite{wave_obs8}, and cardiac arrhythmia~\cite{wave_obs7, arythmia1, arythmia2}. The active vertex model~\cite{selfpropvert1,selfpropvert2}, where activity drives self-propulsion of cells, captures the rigidity transition between solid and fluid states. However, contraction waves can be experimentally observed even when cells' displacements are negligible~\cite{wave_obs3, wave_obs7}. In this context, where cell motility does not play any clear role, it is questionable whether such a motility should be the key ingredient of wave propagation. In fact, the complexity of cellular activity entails the possibility to drive many internal degrees of freedom beyond translational velocity.

Experimental observations of cell area oscillations~\cite{exp_osc_obs2, exp_osc_obs3} and the emergence of phase waves in cardiac tissues~\cite{wave_obs7} motivate novel models. Our working hypothesis is that the collective dynamics in some specific tissues is mostly caused by the cells' deformability. Deformable particles have recently been introduced to study glassy systems~\cite{other_osc1,other_osc2, Brito18prx} and also in the context of nematics~\cite{Yeomans2023, nejad2025, hadjifrangiskou2025}. In general, the broad interest in deformable particles is motivated by their rich emergent behaviors and the potential for designing novel types of actuators~\cite{Manning23prl}.

The introduction of activity as driving the particle sizes is the defining feature of pulsating active matter~\cite{Tjhung2017, Koyano2019, other_osc4, parisi2023, li2024fluidization, zhang, Liu_2024, pineros, banerjee, banerjee2, goth2025, zhu2025, Zhuli2025} with relevance to describe the physics of some living systems~\cite{chiou2016, ishihara2017, Ikeda2019, Staddon-PlosOne2022, boocock2023, Shiladitya2024, tang, koehler2026, desimone2021}. In this class of models, oscillations of particle sizes yield deformation waves akin to experimental observations in biological tissues, despite the absence of self-propulsion. Hydrodynamic theories have strived to describe the large-scale dynamics of such pulsating particles~\cite{banerjee, banerjee2, tang}. In many biological tissues, cells are often anisotropic, so they change not only their sizes but also their overall shape. For instance, cardiomyocytes that form cardiac tissues contract along their long axis~\cite{nitsan2016}. Therefore, it remains to examine how any anisotropic deformation of pulsating particles affects the phenomenology of pulsating active matter.

In this paper, we investigate the role of anisotropy in dense assemblies of pulsating active particles by introducing a drive of some internal degrees of freedom. We extend the model studied in Ref.~\cite{zhang} by considering ellipses that can stabilize nematic order~\cite{DeGennes, Doostmohammadi18natcomm} where the particles' eccentricity depends on an internal phase. Whereas Ref.~\cite{zhang} considers isotropic particles whose internal phase controls only their size, here each particle carries an additional orientational degree of freedom and a phase-dependent eccentricity; this extension reveals how phase synchronisation couples to nematic alignment, distinguishes squeezing from stretching deformations, and leads to nematic order at high density. In Sec.~\ref{sec:micro}, we introduce two models that account for different modes of anisotropic deformation. We build some phase diagrams that report the emergence of various collective states (cycles, arrest, and deformation waves) along with a nematic state at very high densities for a specific deformation mode. In Sec.~\ref{sec:hydro}, we develop a hydrodynamic theory to capture the observed states. The corresponding phase diagram shows qualitative agreement with numerical observations: it allows us to rationalize how anisotropic deformation yields nematic ordering. Finally, we present our conclusions in Sec.~\ref{sec:ccl}.


\section{Collective dynamics of deformable ellipses}\label{sec:micro}

We introduce two distinct models of pulsating deformable ellipses. We outline the repulsive potential acting between the particles, describe the equations of motion that govern their dynamics, and present the order parameters required for a quantitative analysis of the two models.

\subsection{Anisotropic deformation and internal pulsation}

We examine the dynamics of $N$ anisotropic particles in two spatial dimensions. The overdamped dynamics of position $\mathbf{r}_i$ and orientation $\theta_i$ reads
\begin{equation}\label{eq:r}
	\dot{\mathbf{r}}_i = - \mu_r \sum_{j\wedge i} \nabla_iU(a_{ij}) + \sqrt{2 D_r} \boldsymbol{\xi}_i \ ,
	\quad
	\dot{\theta}_i = - \mu_\theta \sum_{j\wedge i} \partial_{\theta_i} U(a_{ij}) + \sqrt{2 D_\theta} \eta_{\theta, i} \ ,
\end{equation}
where $\boldsymbol{\xi_i}$ and $\eta_{\theta,i}$ are uncorrelated Gaussian white noises with zero mean, and correlations given by: $\langle \xi_{i\mu}(t) \xi_{j\nu}(0) \rangle = \delta_{ij} \delta_{\mu\nu}\delta(t)$, and $\langle \eta_{\theta,i}(t) \eta_{\theta,j}(0) \rangle = \delta_{ij} \delta(t)$. The short-ranged repulsive potential is given by $U(a) = a^{12} - 2 a^{6}$ for $a<1$, and $U(a)=0$ otherwise; the sum over $j\wedge i$ thus refers to the interacting neighbors satisfying $a_{ij}<1$. The scaled interparticle distance $a_{ij} = |\mathbf{r}_i - \mathbf{r}_j|/\sigma_{ij}$ is defined in terms of the touching distance $\sigma_{ij}$ between particles. Although $U$ is written as a function of the scalar $a_{ij}$, it depends implicitly on particle orientations through $\sigma_{ij}=\sigma_{ij}(\hat{\mathbf r}_{ij},\theta_i,\theta_j)$. Both $a_{ij}$ and $U(a_{ij})$ are therefore functions of $\hat{\mathbf r}_{ij}$, $\theta_i$, and $\theta_j$.

\begin{figure} 
    \centering 
    \includegraphics[width=.8\textwidth]{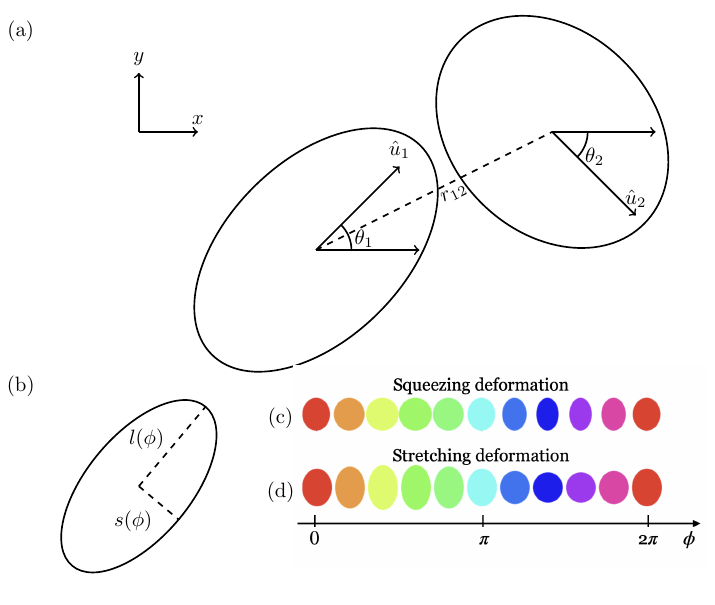}
    \caption{Parametrization of anisotropic deformable particles.
    (a)~Each ellipse has four degrees of freedom: (i)~the position ${\bf r}=(x,y)$ of their center that evolves in two dimensions, the orientation $\theta$ of the long axis $l$, and the internal degree of freedom $\phi$ that controls either the long or short axis, respectively $l(\phi)$ and $s(\phi)$ [Eqs.~\eqref{eq:s} and~\eqref{eq:l}].
    (b-d)~Elliptical deformation as a function of the internal phase $\phi \in [0,2\pi)$. For squeezing (resp.~stretching) deformation, particles are round (resp.~most elliptical) in their maximum size (green, $\phi=\pi/2$), and most elliptical (resp.~round) in their minimum size (dark blue, $\phi=3\pi/2$).
    } 
    \label{fig:ellipse} 
\end{figure}

In all that follows, we focus on particles with elliptical shapes [Fig.~\ref{fig:ellipse}]. Computing the exact analytical expression of $\sigma_{ij}$ for arbitrary shape and orientation of ellipses is both analytically and computationally challenging. Following Ref.~\cite{cleavercare}, we use an analytical approximation for the orientation-dependent contact range of non-equivalent ellipses. Here $\sigma_{ij}$ denotes the centre-to-centre separation at first contact when two ellipses with fixed shapes and orientations are brought together along $\hat{\mathbf r}_{ij}$. The expression is exact for circular particles and provides a standard smooth approximation for the moderate aspect ratios considered here, while avoiding the numerical solution of the exact ellipse-contact problem. Considering two ellipses with short and long axes respectively denoted by $(s_1,s_2)$ and $(l_1,l_2)$, their interparticle direction given by $\hat{\bf r}_{12} = ({\bf r}_1 - {\bf r}_2)/|\mathbf{r}_1 - \mathbf{r}_2|$, and their respective orientations by $\hat{\bf u}_1 = (\cos \theta_1, \sin \theta_1)$ and $\hat{\bf u}_2 = (\cos \theta_2, \sin \theta_2)$, we take the following expression:
\begin{equation}
	\frac{\sigma_{12}}{\sigma_0} =
	\left[ 1 - \frac{\tau\left(\hat{\mathbf{r}}_{12} \cdot \hat{\mathbf{u}}_1\right)^2+\bar \tau\left(\hat{\mathbf{r}}_{12} \cdot \hat{\mathbf{u}}_2\right)^2-2 \tau\bar\tau\left(\hat{\mathbf{r}}_{12} \cdot \hat{\mathbf{u}}_1\right)\left(\hat{\mathbf{r}}_{12} \cdot \hat{\mathbf{u}}_2\right)\left(\hat{\mathbf{u}}_1 \cdot \hat{\mathbf{u}}_2\right)}{1-\tau\bar\tau\left(\hat{\mathbf{u}}_1 \cdot \hat{\mathbf{u}}_2\right)^2} \right]^{-1 / 2} \ ,
\label{eq:sigma1}
\end{equation}
where
\begin{equation}
	\sigma_0^2 = 2(s_1^2+ s_2^2) \ ,
	\quad
	\tau  = \frac{l_1^2-s_1^2}{l_1^2+s_2^2} \ ,
	\quad
	\bar \tau = \frac{l_2^2-s_2^2}{l_2^2+s_1^2} \ .
\end{equation}
We then implement the deformation of a given particle by regarding $(l_i,s_i)$ as dynamical variables. In practice, we consider two deformation rules where either $s_i$ varies at fixed $l_i$, or vice versa. Correspondingly, deformation occurs either by \textit{squeezing} along the short axis, or by \textit{stretching} along the long axis [Figs.~\ref{fig:ellipse}(c-d)]. In the squeezing case, we take $l_i=l_0$ for all particles, while $s_i=s(\phi_i)$ depends on the {\em internal phase} $\phi_i$, which can differ between particles, as 
\begin{equation}\label{eq:s}
	s(\phi) = l_0 \dfrac{1+\lambda \sin{\phi}}{1+\lambda} < l_0 \ ,
\end{equation}
where the deformation constant $\lambda \in (0,1)$ controls the deformation amplitude. In the stretching case, we take instead $s_i=l_0$ for all particles, while $l_i=l(\phi_i)$ varies in terms of $\phi_i$ as 
\begin{equation}\label{eq:l}
	l(\phi) = l_0 \dfrac{1+\lambda \sin \phi}{1-\lambda} > l_0 \ . 
\end{equation}
The definitions in Eqs.~\eqref{eq:s} and~\eqref{eq:l} entail that (i)~squeezing particles are circular at their maximum size ($\phi=\pi/2$) and most elliptical at their minimum size ($\phi=3\pi/2$), whereas (ii)~stretching particles are circular at their minimum size ($\phi=3\pi/2$) and most elliptical at their maximum size ($\phi=\pi/2$); see Figs.~\ref{fig:ellipse}(c-d). Introducing the deformability factor $D$ as the ratio of maximum over minimum area per particle, we find that $D = (1+\lambda)/(1-\lambda)$ in both cases (i-ii), although the maximum and minimum areas differ between cases.

These two deformation rules should be regarded as idealised limiting protocols. The stretching mode, in which deformation is controlled by the long axis, is reminiscent of the predominantly longitudinal contraction of cardiomyocytes, whereas the squeezing mode represents deformation dominated by the short axis. More generally, propagating biochemical activity waves that coordinate tissue growth, such as the ERK waves observed in regenerating osteoblast tissue~\cite{desimone2021}, provide an experimental context in which local deformation and collective wave propagation may be coupled.

Our aim is to explore the nonequilibrium collective effects stemming from an internal pulsation. To this end, inspired by~\cite{zhang, pineros, banerjee, banerjee2}, we introduce an explicit driving term $\omega$ in the dynamics of phases:
\begin{equation}\label{eq:phi}
	\dot{\phi}_i = \omega + \sum_{j\wedge i} \Big[ \varepsilon\sin \left( \phi_j- \phi_i\right)  - \mu_\phi \partial_{\phi_i} U(a_{ij}) \Big] + \sqrt{2 D_\phi}\eta_{\phi, i} \ ,
\end{equation}
where $\eta_{\phi,i}$ is a Gaussian white noise, uncorrelated with $({\boldsymbol\xi}_i,\eta_{\theta,i})$, with zero mean and correlations given by $\langle \eta_{\phi,i}(t) \eta_{\phi,j}(0) \rangle = \delta_{ij} \delta(t)$. The synchronisation amplitude $\varepsilon>0$ promotes alignement between phases of nearby particles, and the synchronisation term has the same range as the potential term deriving from $U$. In the absence of drive and synchronisation ($\omega=0$ and $\varepsilon=0$), the system reduces to an equilibrium assembly of passive deformable particles, provided that $D_r / \mu_r = D_\theta / \mu_\theta = D_\phi/\mu_\phi = T$, where $T$ is the temperature. In all that follows, we fix all the parameters $(D_r, D_\theta, D_\phi, \mu_r, \mu_\phi, \mu_\theta)$ equal to $1$.


\subsection{Phase diagrams: Cycles, arrest, and deformation waves}
 
\begin{figure}
    \centering 
    \includegraphics[width=.9\textwidth, clip=true]{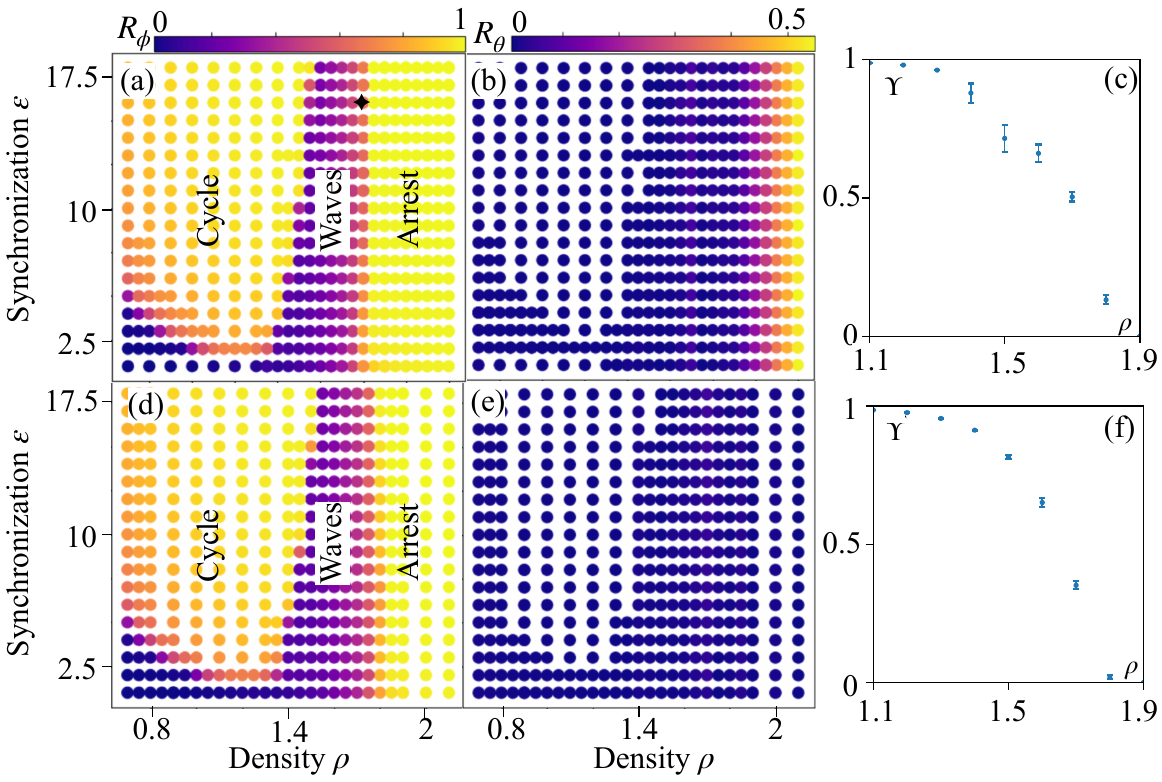}
    \caption{Phase diagrams in terms of synchronisation $R_\phi$ [Eq.~\eqref{eq:Rphi}], nematic order $R_\theta$ [Eq.~\eqref{eq:Rtheta}], and current $\Upsilon$ [Eq.~\eqref{eq:Upsilon}] for squeezing and stretching deformation models (resp.~top and bottom rows). Both models feature arrest, cycle, and wave states in comparable parameter regimes. Only the squeezing deformation model entails a nematic phase ($R_\theta \simeq 1$) at a very high density.
    Parameters: $\omega = 10$, $\lambda = 0.1$, $L = 100$, $l_0 = 0.5$. Panels (c,f): $\varepsilon=10$.
    } 
    \label{fig:phase_diag}
\end{figure}

To quantitatively study the emergence of collective dynamical patterns, we introduce the parameter $R_\phi$ that measures the degree of phase synchronisation among the particles:
\begin{equation}\label{eq:Rphi}
	R_\phi = \frac{1}{N} \bigg\langle\bigg|\sum_{j=1}^N e^{i \phi_j}\bigg|\bigg\rangle \ ,
\end{equation}
the parameter $R_\theta$ that quantifies the degree of nematic alignment between particle orientations:
\begin{equation}\label{eq:Rtheta}
	R_\theta = \frac{1}{N} \bigg\langle\bigg|\sum_{j=1}^N e^{i 2\theta_j}\bigg|\bigg\rangle \ ,
\end{equation}
and the current $\Upsilon$ that quantifies the speed of particle deformation:
\begin{equation}\label{eq:Upsilon}
	\Upsilon = \frac{1}{N\omega} \sum_{j=1}^N \langle\dot\phi_j\rangle .
\end{equation}
Considering particles with either squeezing or stretching deformation [Figs.~\ref{fig:ellipse}(c-d)], we establish phase diagrams by measuring numerically $(R_\phi,R_\theta,\Upsilon)$ for different values of density $\rho_0=N/L^2$, where $L$ denotes the system size, and synchronisation strength $\varepsilon$. Since the current $\Upsilon$ depends only weakly on the synchronisation strength over the range considered, panels (c,f) show representative density cuts at fixed $\varepsilon=10$. In both cases, the system adopts (i)~a synchronized state with vanishing current [$R_\phi\approx 1$, $\Upsilon\approx 0$] at high density, which we refer to as the {\em arrested state}, and (ii)~a synchronized state with maximum current [$R_\phi\approx 1$, $\Upsilon\approx 1$] at moderate density and high synchronisation, which we call the {\em cycling state} [Fig.~\ref{fig:phase_diag}].

\begin{figure}
    \centering 
    \includegraphics[width=.85\textwidth]{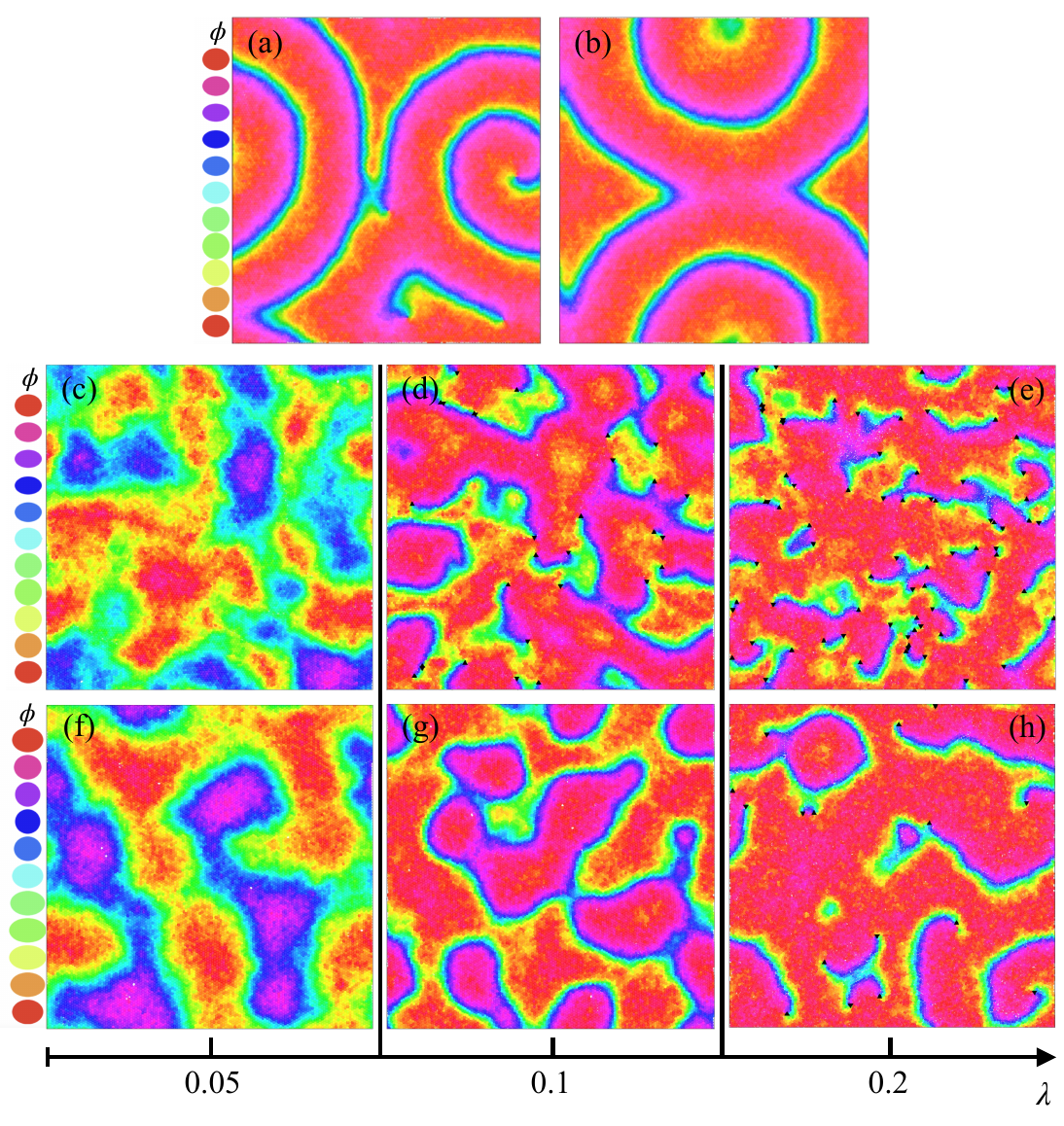} 
    \caption{(a,b)~Patterns observed in the squeezing deformation model for $(\rho,\varepsilon) = (1.8, 16)$, see black diamond in Fig.~\ref{fig:phase_diag}(a).
    (c-h)~Increasing the deformation constant $\lambda$ [Eqs.~(\ref{eq:s}-\ref{eq:l})] makes wavefronts thinner for squeezing and stretching deformation models (resp.~top and bottom rows), and yields topological defects marked with black triangles: upwards for clockwise-rotating defects, and downwards for counterclockwise-rotating defects. At $\lambda = 0.1$, defects form in the squeezing deformation model, but none appear in the stretching deformation model.
    Parameters: $\rho = 1.74$, $\varepsilon = 7.5$, $\omega = 10$, $l_0 = 0.5$.
    }
    \label{fig:defect_lambda}
\end{figure}

The cycling state emerges in the regime where particles have enough space to deform with only moderate repulsion between neighbours. The arrested state arises at higher density due to the repulsion counteracting the drive. Both arrested and cycling states have a homogeneous profile of particle shape. For the density regime in between the arrested and cycling states, there emerges an instability associated with low synchronisation and high current [$R_\phi< 1$, $\Upsilon\approx 1$] that yields deformation waves [Fig.~\ref{fig:defect_lambda}]. This phenomenology is similar to the case of isotropic particles~\cite{zhang}. In addition, we now observe a coupling between the ordering of particle shape and their alignment. At very high density, particles become arrested at an intermediate size fixed by the balance between the drive and the repulsive phase torque, rather than at their absolute minimum size. In this regime, squeezing particles retain an elliptical shape whose steric alignment yields nematic order [$R_\theta>0$, Fig.~\ref{fig:phase_diag}(b)], whereas stretching particles do not develop appreciable nematic order [$R_\theta\approx 0$, Fig.~\ref{fig:phase_diag}(e)].

The deformation waves can spontaneously organize into various patterns. Some of them contain topological defects in the profile of the phase. The charge associated with such defects is defined as $\kappa = \frac{1}{2\pi} \oint d\phi$, where the integration runs over a close loop that contains the defect; it determines whether the defect rotates clockwise ($\kappa=1$) or counterclockwise ($\kappa=-1$). Spiral patterns are associated with a wave that wraps around a defect [Figs.~\ref{fig:defect_lambda}(a)], whereas circular patterns are not associated with any defects [Figs.~\ref{fig:defect_lambda}(b)]. In fact, these two different patterns can emerge for the same values of the control parameters and same initial conditions, as evidence of the metastability of these numerical solutions. The amplitude $\lambda$ of particle deformation [Eqs.~\eqref{eq:s} and~\eqref{eq:l}] significantly influences the defect statistics. Specifically, increasing $\lambda$ leads to more defects for both stretching and squeezing deformations [Fig.~\ref{fig:defect_lambda}], and we observe regimes of $\lambda$ where defects are only present for squeezing deformation.

In short, our models of pulsating deformable ellipses reproduce the three main phases of their isotropic counterpart~\cite{zhang, pineros, manacorda, banerjee}: cycles, arrest, and deformation waves. In addition, we now observe the emergence of a nematically aligned collective state at a very high density for the squeezing deformation [Fig.~\ref{fig:phase_diag}], namely when the smallest particles have an anisotropic shape [Fig.~\ref{fig:ellipse}]. In what follows, we propose a hydrodynamic description to examine the emergence of these various collective states.


\section{Hydrodynamics of synchronisation and nematic order}
\label{sec:hydro}

In this Section, we derive a hydrodynamic description in terms of some fluctuating fields by coarse-graining the microscopic dynamics. First, we propose an effective microscopic dynamics amenable to coarse-graining by simplifying the interaction potential terms. Second, using stochastic calculus~\cite{dean, fodor_lecture}, we derive and close a hierarchy of equations at the hydrodynamic level. The relevant hydrodynamic fields for our model are the density field $\rho$, defined as
\begin{equation}\label{eq:rho}
    \rho(\mathbf{r}, t) = \sum_{j=1}^N \delta(\mathbf{r}-\mathbf{r}_j(t)) \ ,    
\end{equation}
along with the synchronisation $f_\phi$ and nematic $f_\theta$ fields given by
\begin{equation}\label{eq:field}
    f_\phi(\mathbf{r}, t) = \sum_{j=1}^N e^{i\phi_j(t)}\delta(\mathbf{r}-\mathbf{r}_j(t)) \ ,
    \quad
    f_\theta(\mathbf{r}, t) = \sum_{j=1}^N e^{i2\theta_j(t)}\delta(\mathbf{r}-\mathbf{r}_j(t)) \ .
\end{equation} 
By decoupling the displacement of particles from their deformation, we show that the density field $\rho$ relaxes towards a homogeneous profile. The hydrodynamics is then controlled by the coupling between the synchronisation field $f_\phi$ and the nematic field $f_\theta$.


\subsection{Coarse-graining the microscopic dynamics}

To coarse-grain our microscopic model [Eqs.~\eqref{eq:r} and~\eqref{eq:phi}], we start by simplifying the equations of motion. To this end, we adapt the approximation proposed in the case of isotropic particles~\cite{zhang, banerjee} to now account for nematic alignment.

First, we neglect {\it all} interactions in the position dynamics, and we approximate the potential term in the orientation dynamics as
\begin{equation}\label{eq:eff_dyn}
    \dot{\bf r}_i = \sqrt{2D_r} \boldsymbol{\xi}_i  \ ,
    \quad
 		\dot \theta_i =  - \mu_\theta (\partial_\Psi U) (\partial_{\theta_i} \Psi) +\sqrt{2D_\theta} \eta_{\theta, i} \ ,
\end{equation}
Here $U$ is approximated by an effective coarse-grained potential that depends on the local alignment variable $\Psi$, so that the microscopic orientational torque is represented through $\partial_{\theta_i}U\simeq(\partial_\Psi U)(\partial_{\theta_i}\Psi)$. We have introduced the parameter $\Psi$ that quantifies nematic alignment:
\begin{equation}\label{eq:psi}
	\Psi = \frac 1 2 \sum_{j,k=1}^N \cos(2(\theta_j-\theta_k))(1 + \varsigma \nu \sin \phi_j)(1 + \varsigma \nu\sin\phi_k)\delta(\mathbf{r}_j-\mathbf{r}_k) \ ,
\end{equation}
The potential $U$ should decrease with the alignment parameter $\Psi$, so that $\partial_\Psi U<0$. The coupling constant $\nu$ accounts for the fact that the potential term $\partial_{\theta_i} U$ depends not only on the orientation $\theta_i$, but also on the shape of particles through the phases $\phi_i$. Specifically, at fixed $\theta_i$, we expect that $\partial_{\theta_i} U$ vanishes when particles are isotropic, and increases with anisotropy. This trend is captured by taking $\varsigma=-1$ for squeezing deformation, and $\varsigma=1$ for stretching deformation. In what follows, we consider $\nu$ as a small parameter, in line with taking a small amplitude of particle deformation [Eqs.~\eqref{eq:s} and~\eqref{eq:l}]:
\begin{equation}
    \Psi = \frac 1 2 \sum_{j,k=1}^N \cos(2(\theta_j-\theta_k)) \Big[ 1 + \varsigma \nu (\sin \phi_j + \sin\phi_k) + o(\nu) \Big] \delta(\mathbf{r}_j-\mathbf{r}_k) \ ,
\end{equation}
where we have truncated $\Psi$ to first order in $\nu$.

Second, we approximate the potential terms in the phase dynamics as
\begin{equation}\label{eq:eff_dyn_bis}
	\dot \phi_i = \omega - \mu_\phi (\partial_\Psi U)(\partial_{\phi_i}\Psi )  + \sum_{j=1}^N \Big[ \varepsilon \sin(\phi_j-\phi_i) - \mu_\phi (\partial_\varphi U)(\partial_{\phi_j}\varphi) \Big] \delta(\mathbf{r}_i-\mathbf{r}_j) + \sqrt{2D_\phi} \eta_{\phi, i} \ ,
\end{equation}
where we have introduced the packing fraction 
\begin{equation}
	\varphi = \dfrac{\pi}{L^2} \sum_{j=1}^N l(\phi_j) s(\phi_j) \ .
\end{equation}
The potential $U$ should increase with the packing fraction $\varphi$, so that $\partial_\varphi U>0$. The long and short axes $(l,s)$ are defined differently for stretching [Eq.~\eqref{eq:s}] and squeezing [Eq.~\eqref{eq:l}] deformations, and $L$ refers to the system size. We expect that the assumptions underlying the effective dynamics [Eqs.~\eqref{eq:eff_dyn} and~\eqref{eq:eff_dyn_bis}] hold in the dense regime with a strong overlap between particles.

The time evolution of the empirical distribution
\begin{equation}
	f(\mathbf{r}, \phi, \theta, t) = \sum_{i=1}^N \hat{f}_i(\mathbf{r},\theta, \phi, t) \ ,
	\quad
	\hat{f}_i(\mathbf{r},\theta, \phi, t) = \delta({\bf r}-{\bf r}_i(t))\delta(\theta-\theta_i(t))\delta(\phi-\phi_i(t)) \ ,
\end{equation}
can be written using stochastic calculus~\cite{dean, fodor_lecture} as
\begin{equation}\label{eq:te_f}
	\partial_t f =\sum_{i=1}^N \Big[ \dot {\bf r}_i \cdot \partial_{\mathbf{r}_i} + \dot \theta_i \partial_{\theta_i} + \dot \phi_i \partial_{\phi_i} + D_r \partial_{\mathbf{r}_i\mathbf{r}_i}^2 + D_\theta \partial_{\theta_i \theta_i}^2 + D_\phi \partial_{\phi_i \phi_i}^2 \Big] \hat{f}_i \ .
\end{equation}
Substituting the effective microscopic dynamics [Eqs.~\eqref{eq:eff_dyn} and~\eqref{eq:eff_dyn_bis}] into Eq~\eqref{eq:te_f}, and introducing the mode decomposition
\begin{equation}
    f_{n,m} (\mathbf{r}, t) = \int d\phi d\theta e^{i(n\phi+2m\theta)} f(\mathbf{r}, \phi, \theta, t) = \sum_{j=1}^N e^{i(n\phi_j(t)+2m\theta_j(t))}\delta(\mathbf{r}-\mathbf{r}_j(t)) \ ,
    \label{eq:ft}
\end{equation}
we obtain an infinite hierarchy of equations:
\begin{equation}
 	\label{eq:fnm}
	\begin{aligned}
		\partial_t f_{n,m} &= (D_r \partial_{{\bf r}{\bf r}}^2 - n^2 D_\phi - 4 m^2 D_\theta + in\omega -in c_{\phi,1}Re[f_{1,0}])f_{n,m}
    \\
    &\quad + \frac{n\varepsilon}{2} (f_{1,0}f_{n-1,m}-f_{1,0}^* f_{n+1,m})
		\\
    &\quad -i \varsigma \nu  n c_{\phi,2} \Big[ f_{0,1}(f_{n-1,m-1}+f_{n+1,m-1}) +  f_{0,1}^* (f_{n-1,m+1}+f_{n+1,m+1}) \Big]
    \\
    &\quad + 2i\varsigma  m c_\theta \nu \Big[ f_{n,m-1}(f_{1,1}-f_{1,-1}^*)+  f_{n,m+1}(f_{1,1}^*-f_{1,-1}) \Big]
    \\
    &\quad - 2i \varsigma m c_\theta \nu \Big[ f_{0,1}^* (f_{n+1,m+1}-f_{n-1,m+1}) + f_{0,1} (f_{n-1,m-1}-f_{n+1,m-1}) \Big]
    \\
    &\quad + 2m c_\theta ( f_{0,1}^* f_{n,m+1} - f_{0,1} f_{n,m-1} ) + \Lambda_{n,m} \ ,
  \end{aligned}
\end{equation}
where $f_{n,m}^*=f_{-n,-m}$ refers to the complex conjugate of $f_{n,m}$, and
\begin{equation}
	c_{\phi,1} = \mu_\phi (\partial_\varphi U)\dfrac{\lambda\pi l_0}{(1+\varsigma\lambda)L^2} > 0 \ ,
	\quad
	c_\theta = \frac{\mu_\theta \lambda}{4} (\partial_\Psi U) < 0 \ ,
	\quad
	c_{\phi,2} = \frac{\mu_\phi \lambda}{4} (\partial_\Psi U) < 0 \ .
\end{equation}
The noise term reads
\begin{equation}\label{eq:noise}
	\Lambda_{n,m} = \sum_{j=1}^N e^{i(n\phi_j+2m\theta_j)} \Big[ \sqrt{2D_r} {\boldsymbol\xi}_j \cdot \partial_{\bf r} + in\sqrt{2D_\theta} \eta_{\theta, j} + 2im \sqrt{2D_\phi} \eta_{\phi, j} \Big]  \delta({\bf r}-{\bf r}_j) \ .
\end{equation}
For $(n,m)=(0,0)$, the dynamics of the density $f_{0,0}=\rho$ [Eq.~\eqref{eq:rho}] decouples from higher-order modes: $\partial_t f_{0,0} = D_r \partial^2_{{\bf r}{\bf r}} f_{0,0}$, where we have neglected the noise contribution. It follows that the density field relaxes to the homogeneous profile $\rho=\rho_0=N/L^2$.

The fields of interest $(f_\phi,f_\theta)$ [Eq.~\eqref{eq:field}] respectively coincide with the modes $(f_{1,0}, f_{0,1})$ [Eq.~\eqref{eq:ft}]. To close the hydrodynamics in terms of these modes, we make a scaling ansatz which leads to the adiabiatic elimination of higher modes~\cite{zhang}: $f_{n,m} \sim \zeta^{|n|+|m|}$ and $\partial_t\sim \partial_{rr}\sim \zeta^2$, where $\zeta\ll 1$. This ansatz is justified close to the disordered state, where we expect the distribution of phases and orientations to be almost uniform~\cite{Bertin09jphysa}. It follows that $O(\zeta^3)$ terms are subleading in the evolution of the second-order modes $(f_{2,0}, f_{0,2}, f_{1,1}, f_{1,-1})$, yielding
\begin{equation}\label{eq:f11}
\begin{aligned}
	f_{2,0} &= \frac{\varepsilon f_{10}^2}{2(2D_\phi -i\omega)} \ ,
	\quad
	f_{0,2} = - \frac{c_\theta}{D_\theta}f_{01}^2 \ ,
	\\
	f_{1,1} &= \frac{\varepsilon f_{01}f_{10}}{2(D_\phi + D_\theta - i \omega)}\ ,
	\quad
	f_{1,-1} = \frac{\varepsilon f_{10}f_{01}^*}{2(D_\phi + D_\theta -i\omega)} \ .
\end{aligned}
\end{equation}
Substituting Eq.~\eqref{eq:f11} in the evolution of $(f_{1,0}, f_{0,1}) = (f_\phi, f_\theta)$ [Eq.~\eqref{eq:fnm}], we get
\begin{equation}\label{eq:HD}
	\begin{aligned}
    \partial_t f_{\phi} &= (D_r \partial_{{\bf r}{\bf r}}^2 - D_\phi + \varepsilon \rho_0/2 +i\omega - ic_{\phi,1} {\rm Re}[f_{\phi}] )f_{\phi}
    \\ 
    &\quad -\dfrac{\varepsilon^2}{4(2D_\phi - i\omega)} |f_{\phi}|^2f_{\phi} -i2\varsigma \nu c_{\phi,2}|f_{\theta}|^2 + \sqrt{2\rho_0 D_\phi} \eta_\phi \ ,
    \\
    \partial_t f_{\theta} &= (D_r\partial_{{\bf r}{\bf r}}^2-D_\theta -2c_\theta \rho_0)f_{\theta} - 2\dfrac{c_\theta^2}{D_\theta}|f_{\theta}|^2f_{\theta} -2\varsigma \nu  c_\theta f_{\theta}{\rm Im}[f_{\phi}]
    \\
    &\quad -\varsigma \varepsilon c_\theta \nu  \rho_0f_{\theta} \dfrac{(D_\phi + D_\theta ) {\rm Im}[f_{\phi}]+\omega {\rm Re}[f_{\phi}]}{(D_\phi+D_\theta)^2 + \omega^2} + \sqrt{8\rho_0 D_\theta} \eta_\theta \ ,
\end{aligned}
\end{equation}
where $(\eta_\phi,\eta_\theta)$ are some uncorrelated Gaussian white noises with zero mean and unit correlation [\ref{app:noise}].

Therefore, our coarse-graining yields a closed hydrodynamics [Eq.~\eqref{eq:HD}] for the synchronisation and nematic fields $(f_\phi,f_\theta)$ [Eq.~\eqref{eq:field}]. This hydrodynamics is invariant under an arbitrary rotation of the nematic field ($f_\theta\to f_\theta e^{i\psi}$ for an arbitrary real number $\psi$), yet a similar rotation of the synchronisation field ($f_\phi\to f_\phi e^{i\psi}$) is not invariant. Such a broken invariance has already been identified in the case of pulsating isotropic particles~\cite{zhang, banerjee, banerjee2}. The noise-dominated regime is represented in Eq.~\eqref{eq:HD}: when $D_\phi$ is large compared with the interaction-induced phase torque, the damping term $-D_\phi f_\phi$ suppresses phase order, yielding the disorder regime labelled in Fig.~\ref{fig:theo_pd_sqst}(a,b). In the position dynamics, neglecting the interaction force leaves translational diffusion as the dominant contribution, so that the density relaxes towards a homogeneous profile. For $\nu=0$, the phase and nematic sectors decouple. The phase sector is qualitatively analogous to the isotropic-particle hydrodynamics, but an exact reduction additionally requires removing the independent nematic sector and retaining the contribution of the second phase harmonic $f_{2,0}$ to the packing-fraction torque, which appears as an additional imaginary-part term after closure~\cite{zhang}.


\subsection{Phase diagram: Weak coupling of alignment and deformation}

We consider a perturbative approach of the hydrodynamics [Eq.~\eqref{eq:HD}] in terms of the coupling parameter $\nu$: we first solve the case $\nu = 0$, and then set $\nu$ as a small parameter to determine corrections. We examine homogeneous field realizations, decompose them as $(f_\phi, f_\theta) = (R_\phi e^{i\Phi}, R_\theta e^{i2\Theta})$, and substitute these expressions into the hydrodynamics [Eq.~\eqref{eq:HD}]. Note that $\Theta$ is undetermined, since the direction of $f_\theta$ is invariant under rotation.

\begin{figure} 
	\centering 
	\includegraphics[width=1\textwidth]{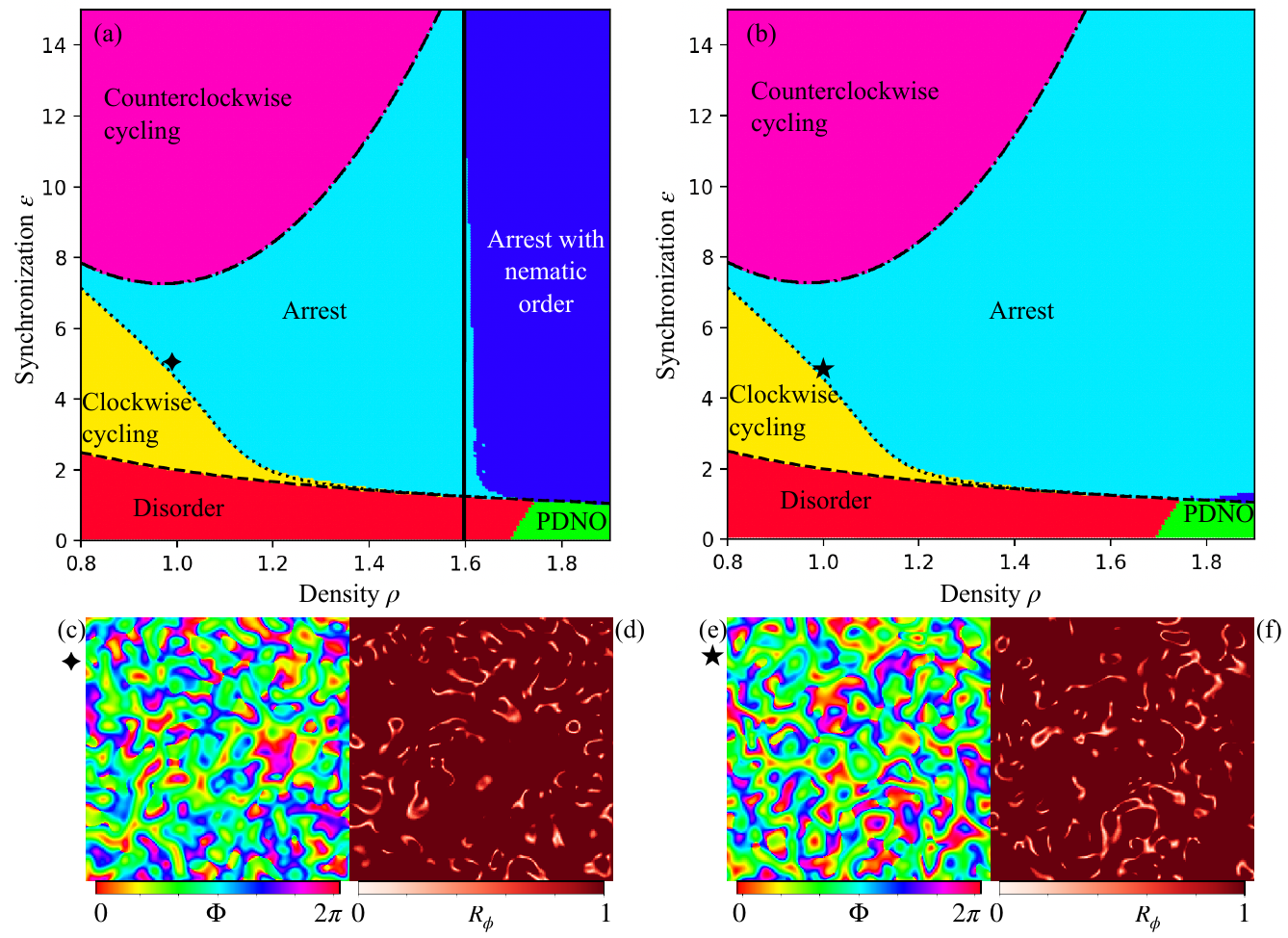} 
	\caption{Hydrodynamic phase diagrams for the cases of (a)~squeezing and (b)~stretching deformation models. In panels (a,b), the lines denote the analytical phase boundaries, while the coloured regions are determined by numerical integration of the homogeneous hydrodynamic equations. Black lines delinate the regions of existence of disordered (PDNO refers to phase disorder with nematic order), arrested, and cycling (clockwise and counter-clockwise) states at $\nu=0$; see dashed [Eq.~\eqref{eq:dashed}], dotted and dash-dotted [Eq.~\eqref{eq:dotted}], and solid [Eq.~\eqref{eq:solid}] lines. To first order in $\nu$, the linear stability analysis of the fixed-points reveals that nematic order emerges at a smaller density for squeezing than for stretching deformation.
	Profile of the synchronisation field $f_\phi = R_\phi e^{i\Phi}$ [Eq.~\eqref{eq:field}] obtained from numerical integration of hydrodynamics [Eq.~\eqref{eq:HD}] for (c,d)~the squeezing deformation model ($\varsigma = -1$), and (e,f)~the stretching deformation model ($\varsigma = 1$).
	 Parameters: $\rho_0 = 1$, $\varepsilon = 6$, $\omega = 10$, $L = 1$, $dx = 1/256$, $\nu = 0.2$, $c_{\phi,2} = c_\theta = -0.01 \rho_0^6$, $c_{\phi,1} = \rho_0^6$.
	} 
\label{fig:theo_pd_sqst}
\end{figure}

The fixed-point solutions $(R_{\phi,0}, \Phi_0, R_{\theta,0})$ at $\nu=0$ obey
\begin{equation}\label{eq:syst_homo}
  \begin{aligned}
		0 &= \left(\frac{\varepsilon\rho_0}{2}-D_\phi\right) R_{\phi,0} - \dfrac{D_\phi X(\varepsilon,D_\phi,\omega)}{2}R_{\phi,0}^3  \ ,
		\\
		0 &= \omega R_{\phi,0} - c_{\phi,1}\cos\Phi_0 R_{\phi,0}^2 - \dfrac{\omega X(\varepsilon,D_\phi,\omega)}{4}R_{\phi,0}^3 \ ,
		\\
		0 &= (2c_\theta \rho_0\color{black} + D_\theta)R_{\theta,0} + 2\dfrac{c_\theta^2}{D_\theta}R_{\theta,0}^3 \ ,
  \end{aligned}
\end{equation}
where
\begin{equation}
	X(\varepsilon,D_\phi,\omega) = \dfrac{\varepsilon^2}{4D_\phi^2+\omega^2} .
\end{equation}
These solutions lead to distinguish four different states:
\begin{itemize}
    \item disorder: $(R_{\phi,0},R_{\theta,0})=(0,0)$,
    \item phase disorder and nematic order: $(R_{\phi,0},R_{\theta,0})=(0,\bar R_{\theta, 0})$,
    \item arrested state and nematic disorder: $(R_{\phi,0}, \Phi_0, 0 ) = (\bar R_{\phi,0}, \bar \Phi_0, 0)$,
    \item arrested state and nematic order: $(R_{\phi,0}, \Phi_0,R_{\theta,0})=(\bar R_{\phi,0}, \bar \Phi_0,\bar R_{\theta, 0})$ ,
\end{itemize}
in terms of
\begin{equation}\label{eq:FP0}
	\bar R_{\phi,0} = \sqrt{\dfrac{\varepsilon \rho_0 - 2D_\phi}{D_\phi X(\varepsilon,D_\phi,\omega)}} \ ,
	\
	\bar R_{\theta,0} = \sqrt{\dfrac{D_\theta}{c_\theta}\left(\rho_0 - \frac{D_\theta}{2c_\theta}\right)} \ ,
	\
	\cos\bar\Phi_0 = \dfrac{1}{c_{\phi,1}} \dfrac{\omega}{\bar R_{\phi,0} }\left( \frac{3}{2} - \dfrac{\varepsilon\rho_0}{D_\phi}\right)\ .
\end{equation}
The condition $\bar R_{\phi,0}^2>0$ determines the critical line for the existence of ordered states:
\begin{equation}\label{eq:dashed}
    \varepsilon\rho_0  = \dfrac{2D_\phi}{\varepsilon} \ ,
\end{equation}
where the cycling and arrested states are separated by the following condition:
\begin{equation}\label{eq:dotted}
	\cos \bar\Phi_0 = \pm 1 ,
\end{equation}
while $\bar R_{\theta,0}^2>0$ yields the critical line for the existence of nematic order:
\begin{equation}\label{eq:solid}
    \rho_0 =  \frac{D_\theta}{2c_\theta} \ ,
\end{equation}
as reported in Fig.~\ref{fig:theo_pd_sqst}.

To study the stability of the fixed-point solutions at $\nu=0$ [Eq.~\eqref{eq:syst_homo}], we consider the dynamics of the linear perturbation
\begin{equation}
	\delta\mathbf X_0(q,t) = \int d{\bf r} [ R_\phi - R_{\phi,0}, \Phi - \Phi_0 , R_\theta - R_{\theta,0} ] e^{i {\bf q}\cdot{\bf r}} \ ,
	\quad
	\delta\dot{\mathbf X}_0 = M_0 \delta\mathbf X_0 \ ,
\end{equation}
where the stability matrix $M_0$ reads
\begin{equation}\label{eq:m0}
    M_0 =
    \begin{pmatrix}
       m_{0,11} - D_r q^2 & 0  & 0\\
       m_{0,21} & m_{0,22} - D_r q^2 & 0\\
       0 & 0 &  m_{0,33} - D_r q^2
    \end{pmatrix} ,
\end{equation}
in terms of
\begin{equation}
    \begin{aligned}
        m_{0,11} &= \frac{\varepsilon \rho_0}{2} - D_\phi - \frac{3 D_\phi X(\varepsilon,D_\phi,\omega)}{2} R_{\phi,0} \ ,
        \\
        m_{0,21} &= \dfrac{\omega}{R_{\phi,0}} - \dfrac{3\omega X(\varepsilon,D_\phi,\omega)}{4} R_{\phi,0}- 2c_{\phi,1}\cos\Phi_0 \ ,
        \\
        m_{0,22} &= R_{\phi,0}c_{\phi,1}\sin \Phi_0 \ ,
        \quad
        m_{0,33} = - D_\theta - 2 c_\theta \rho_0 - \frac{6 c_\theta^2}{D_\theta}  R_{\theta,0}^2 \ .
    \end{aligned}
\end{equation}
The eigenvalues $\chi_0(q)$ of $M_0$ take the form $\chi_0 (q) = \chi_0 (0) - D_r q^2$, where $\chi_0 (0) < 0$ in the regime of existence of the fixed-point solutions, so that these solutions are linearly stable at all $q$. In short, when alignment and deformation decouple ($\nu=0$), the phase diagram is the same for both the squeezing and stretching deformation models: the only difference with the hydrodynamics of isotropic pulsating particles~\cite{zhang} is the emergence of nematic order at high density.

We now examine the fixed-point solutions $(R_{\phi,1}, \Phi_1, R_{\theta,1})$ that are the first-order corrections to the case $\nu=0$:
\begin{equation}
    \begin{aligned}
		R_{\phi,1} &= \dfrac{2\varsigma c_{\phi,2}R_{\theta,0}^2 \sin\Phi_0 }{\dfrac{\varepsilon \rho_0}{2} - D_\phi - \dfrac{3\omega X(\varepsilon,D_\phi,\omega)}{4} R_{\phi,0}^2} \ ,
		\\
		\Phi_1 &= \dfrac{R_{\phi,1} \left( 2c_{\phi,1} R_{\phi,0} \cos\Phi_0 - \omega + \dfrac{3\omega X(\varepsilon,D_\phi,\omega)}{4} R_{\phi,0}^2 \right) + 2\varsigma c_{\phi,2} R_{\theta,0}^2 \cos\Phi_0}{c_{\phi,1} R_{\phi,0}^2 \sin\Phi_0} \ ,
		\\
		R_{\theta,1} &= -\dfrac{\varsigma  D_\theta}{2c_\theta} \dfrac{R_{\phi,0}}{R_{\theta,0}} \sin\Phi_0 - \dfrac{\varsigma  D_\theta \varepsilon \rho_0}{2c_\theta} \dfrac{R_{\phi,0}}{R_{\theta,0}} \dfrac{(D_\phi +D_\theta)\sin\Phi_0+\omega \cos \Phi_0}{(D_\phi + D_\theta)^2+\omega^2} \ .
    \end{aligned}
\end{equation}
The dynamics of the corresponding linear perturbation is given by
\begin{equation}
\begin{aligned}
	\delta\mathbf X_1(q,t) &= \int d{\bf r} [ R_\phi - R_{\phi,0} - \nu R_{\phi,1}, \Phi - \Phi_0 - \nu \Phi_1 , R_\theta - R_{\theta,0} - \nu R_{\theta,1} ] e^{i {\bf q}\cdot{\bf r}} \ ,
	\\
	\delta\dot{\mathbf X}_1 &= (M_0 + \nu M_1) \delta\mathbf X_1 \ ,
\end{aligned}
\end{equation}
where $M_0$ is defined in Eq.~\eqref{eq:m0}, and the components of $M_1$ read
\begin{equation}
\begin{aligned}
	M_{1,11} &= -3 D_\phi X(\varepsilon, D_\phi, \omega) R_{\phi,0} R_{\phi,1} \ ,
	\\
	M_{1,12} &= -2\varsigma c_{\phi,2} { R_{\theta,0}}^2\cos \Phi_0 \ ,
	\\
	M_{1,13} &= -4\varsigma c_{\phi,2} { R_{\theta,0}}\sin \Phi_0 \ ,
\end{aligned}
\end{equation}    
and
\begin{equation}
\begin{aligned}
M_{1,21} &= -\omega  R_{\phi,1}\dfrac{\varepsilon^2 (2D_\phi-3\varepsilon\rho_0)}{4(2D_\phi-\varepsilon \rho_0)(4D_\phi^2+\omega^2)}-2c_{\phi,1} \Phi_1 \sin \Phi_0 \ ,
\\
M_{1,22} &= c_{\phi,1} R_{\phi,1}\sin \Phi_0 + c_{\phi,1} R_{\phi,0} \Phi_1 \cos \Phi_0 + 2\varsigma c_{\phi,2}\dfrac{ R_{\theta,0}^2}{ R_{\phi,0}}\sin \Phi_0 \ ,
\\
M_{1,23} &= -4\varsigma c_{\phi,2}\dfrac{ R_{\theta,0}}{ R_{\phi,0}}\cos \Phi_0 \ ,
\end{aligned}
\end{equation}    
and
\begin{equation}
\begin{aligned}
    M_{1,31} &= -2\varsigma c_\theta  R_{\theta,0} \sin \Phi_0 - \varsigma \varepsilon c_\theta \rho_0 \dfrac{(D_\phi+D_\theta)\sin \Phi_0+\omega \cos \Phi_0}{(D_\phi+D_\theta)^2+\omega^2} R_{\theta,0} \ ,
    \\
    M_{1,32} &= -2\varsigma c_\theta  R_{\theta,0} R_{\phi,0}\cos \Phi_0 + \varsigma \varepsilon c_\theta \rho_0 \dfrac{(D_\phi+D_\theta)\cos \Phi_0-\omega \sin \Phi_0}{(D_\phi+D_\theta)^2+\omega^2} R_{\theta,0} R_{\phi,0} \ ,
    \\
    M_{1,33} &= -\dfrac{3c_\theta^2}{D_\theta} R_{\theta,0} R_{\theta,1}-2\varsigma c_\theta  R_{\phi,0}\sin \Phi_0 - \varsigma \varepsilon c_\theta \rho_0 \dfrac{(D_\phi+D_\theta)\sin \Phi_0 + \omega \cos \Phi_0}{(D_\phi+D_\theta)^2+\omega^2} R_{\phi,0} \ .
\end{aligned}
\end{equation}
The eigenvalues $\chi$ of $M = M_0 + \nu M_1$ can be written perturbatively as~\cite{horn2012, stewart1990}
\begin{equation}\label{eq:chi}
	\chi_\alpha = \chi_{0,\alpha} + \nu \ \dfrac{({\bf y}^* \cdot M_1 \cdot {\bf x})_\alpha}{{\bf y}^* \cdot {\bf x}} + o(\nu) \ ,
\end{equation}
where $\chi_0$ are the eigenvalues of $M_0$, $({\bf y},{\bf x})$ respectively denote the right and left eigenvectors of $M_0$, and ${\bf y}^*$ is the conjugate transpose of $\bf y$.

Based on the eigenvalues of the stability matrix [Eq.~\eqref{eq:chi}], we deduce the phase diagram outlining the stability regions of the homogeneous stationary solutions in both the squeezing ($(\varsigma = -1$) and stretching ($\varsigma = 1$) deformation models. We define counter-clockwise (clockwise) cycling by $\langle\dot\Phi\rangle>0$ ($\langle\dot\Phi\rangle<0$), corresponding respectively to positive and negative hydrodynamic phase current. We find that both models exhibit (i)~two cycling states (clockwise and counterclockwise), (ii)~a disordered state, and (iii)~an arrested state [Figs.~\ref{fig:theo_pd_sqst}(a-b)]. We observe that nematic order emerges at a smaller density for squeezing than for stretching deformation model, showing that our perturbative treatment in $\nu$ effectively captures the key distinction between these models. Overall, the features of the hydrodynamic phase diagram are in qualitative agreement with the particle-based results [Fig.~\ref{fig:phase_diag}], except for the clockwise cycling that has no counterpart in the microscopic model, and that was also reported in the hydrodynamics of isotropic pulsating particles~\cite{zhang}.

In the presence of noise, the hydrodynamics leads to the emergence of dynamical patterns [Figs.~\ref{fig:theo_pd_sqst}(c-f)] in a regime of parameters between the cycling and arrested states. These patterns take the form of propagating waves associated with defects in the profile of the synchronisation field $f_{\phi} = R_\phi e^{i\Phi}$ without any noticeable difference between the stretching and squeezing deformation models. Such a phenomenology is in qualitative agreement with the particle-based simulations [Fig.~\ref{fig:defect_lambda}], and also similar to the patterns in the hydrodynamics of isotropic pulsating particles~\cite{zhang}. In particular, patterns appear at hydrodynamic level despite neglecting any correlation between the synchronisation and density fields, since we assume $\rho$ to be constant.


\section{Discussion}\label{sec:ccl}

We have investigated the impact of anisotropy on the collective dynamics of active pulsating ellipses. Considering two types of anisotropy, either squeezing or stretching, we report collective states analogous to the case of isotropic pulsating particles~\cite{zhang, pineros}, including circular and spiral waves. In the squeezing case, where particles are elliptical at minimum size, nematic order emerges at high densities, in contrast with the isotropic and stretching cases. We have rationalized our numerical results of molecular dynamics through a hydrodynamic description that we obtain by coarse-graining of the microscopic equations. The corresponding hydrodynamic equations extend the ones derived in~\cite{zhang, banerjee, banerjee2}, by now taking into account the interplay between phase synchronisation and nematic order.

Overall, our results contribute to a better understanding of how shape anisotropy affects the collective dynamics in pulsating active systems. Our findings could be particularly relevant for modeling biological tissues where cells often exhibit both area and shape changes, as seen in systems like cardiac tissue and epithelial layers~\cite{wave_obs7, arythmia1, arythmia2, tang}. Future works, either experimental or theoretical, could analyze more quantitatively the role of the individual deformation parameter $\lambda$ [Eqs.~\eqref{eq:s} and~\eqref{eq:l}] in controlling defect formation, for instance to prevent the formation of spiral waves in cardiac tissues. A possible direction for future work is to vary independently the translational, rotational, and phase diffusivities and mobilities, in order to determine how the ratios $D_r/\mu_r$, $D_\theta/\mu_\theta$, and $D_\phi/\mu_\phi$ modify the balance between fluctuations, mechanical relaxation, synchronisation, and drive. An interesting extension would be to consider area-preserving deformations, in which the particle aspect ratio varies while its area remains constant. In this case, the direct coupling between phase and packing fraction would vanish, while phase-dependent excluded-volume interactions could still couple pulsation to density and nematic order. Such a protocol could therefore help disentangle the respective roles of area and shape variations in the collective dynamics. 

\section*{Acknowledgements}

This project has received funding from the European Union’s Horizon Europe research and innovation programme under the Marie Sk\l{}odowska-Curie grant agreement No 101056825 (NewGenActive), and from the Luxembourg National Research Fund (FNR), grant references 17962137 and 14389168. A.M. acknowledges financial support from Grant No. 2022HNW5YL MOCA funded by the Ministero dell’Università e della Ricerca PRIN2022 program.\\




\appendix
\section{Hydrodynamic noise}\label{app:noise}

In this Appendix, we compute the statistics of the noise terms $\Lambda_{10}$ and $\Lambda_{01}$ [Eq.~\eqref{eq:noise}], defined as
\begin{equation}\label{eq:noise1}
	\Lambda_{1,0} = i\sum_{j=1}^N e^{i\phi_j}\sqrt{2D_\phi}\,\eta_{\phi,j}\delta(\mathbf{r}-\mathbf{r}_j) \ ,
	\quad
  \Lambda_{0,1}= i2\sum_{j=1}^N e^{i2\theta_j}\sqrt{2D_\theta}\,\eta_{\theta,j}\delta(\mathbf{r}-\mathbf{r}_j) \ .
\end{equation}
Assumming the scaling $\partial_r\sim\zeta$, where $\zeta\ll 1$, we have neglected the contribution from the conserved noise in the expression of $(\Lambda_{1,0},\Lambda_{0,1})$ [Eq.~\eqref{eq:noise}]. Since $\eta_{\phi,j}$ and $\eta_{\theta,j}$ are Gaussian noises, $\Lambda_{01}$ and $\Lambda_{10}$ are also Gaussian too, and their correlations read
\begin{equation}
  \begin{aligned}
    \left\langle \Lambda_{1,0}(\mathbf{r},t) \Lambda^*_{1,0}(\mathbf{r}',t')\right\rangle &= 2 D_\phi f_{0,0}(\mathbf{r},t) \delta(\mathbf{r}-\mathbf{r}') \delta(t-t') \ ,
    \\
    \left\langle \Lambda_{1,0}(\mathbf{r},t) \Lambda_{1,0}(\mathbf{r}',t')\right\rangle &= - 2 D_\phi f_{2,0}(\mathbf{r},t) \delta(\mathbf{r}-\mathbf{r}') \delta(t-t') \ ,
    \\
    \left\langle \Lambda_{0,1}(\mathbf{r},t) \Lambda^*_{0,1}(\mathbf{r}',t')\right\rangle &= 8 D_\theta f_{0,0}(\mathbf{r},t) \delta(\mathbf{r}-\mathbf{r}') \delta(t-t') \ ,
    \\
    \left\langle \Lambda_{0,1}(\mathbf{r},t) \Lambda_{0,1}(\mathbf{r}',t')\right\rangle &= - 8 D_\theta f_{0,2}(\mathbf{r},t) \delta(\mathbf{r}-\mathbf{r}') \delta(t-t') \ .
  \end{aligned}
\end{equation}
The second-order modes $f_{2,0}$ and $f_{0,2}$ are of higher order in $\zeta$ compared with $f_{00}$, so that we subsequently neglect them. Therefore, the real and imaginary parts are independent for both $\Lambda_{10}$ and $\Lambda_{01}$ a to leading order. Finally, we can write
\begin{equation}
	\Lambda_{1,0} = \sqrt{2\rho_0 D_\phi} \eta_\phi \ ,
	\quad
	\Lambda_{0,1} = \sqrt{8\rho_0 D_\theta} \eta_\theta \ ,
\end{equation}
where $(\eta_\phi, \eta_\theta)$ are some uncorrellated Gaussian white noises with zero mean and correlations given by
\begin{equation}
	\langle \eta_\phi({\bf r},t) \eta_\phi^*({\bf r'},t') \rangle = \langle \eta_\theta({\bf r},t) \eta_\theta^*({\bf r'},t') \rangle = \delta({\bf r}-{\bf r}') \delta(t-t') \ ,
\end{equation}
where we have used that $f_{00}$ relaxes to the homogeneous profile $\rho_0$.


\section*{References}

\bibliographystyle{iopart-num}
\bibliography{library}

\end{document}